\documentclass[12pt,a4paper,english,superscriptaddress,aps,nofootinbib]{revtex4}
\usepackage[utf8]{inputenc}
\usepackage[T1]{fontenc}
\usepackage{amsmath}
\usepackage{amssymb}
\usepackage{graphicx}
\makeatletter
\usepackage{babel}
\usepackage[active]{srcltx}
\usepackage{graphicx,color}
\usepackage{subfigure}
\usepackage{changebar}
\usepackage{upgreek}
\usepackage{hyperref}
\hypersetup{colorlinks=true,urlcolor= blue,citecolor=blue,linkcolor= blue}
\usepackage{braket}
\usepackage{dsfont}
\usepackage{mathtools}
\usepackage{slashed}
\usepackage{empheq}
\usepackage{tikz}
\usepackage{tikz-feynman}
\usepackage{multirow}
\usetikzlibrary{decorations.pathmorphing}

\usepackage{tikz}
\usepackage{tikz-feynman}
\tikzfeynmanset{compat=1.1.0}

\usepackage{graphicx}
\usepackage{amsfonts}
\usepackage{amsmath}
\newcommand{\be}{\begin{equation}}
	\newcommand{\ee}{\end{equation}}
\newcommand{\bea}{\begin{eqnarray}}
	\newcommand{\eea}{\end{eqnarray}}

\begin{document}

\title{Magnetized vortex in three-dimensional $\mathrm{f(R)}$ gravity}

\author{F. C. E. Lima}
\email[]{E-mail: cleiton.estevao@ufabc.edu.br; cleiton.estevao@ufabc.edu.br}
\affiliation{Departamento de F\'{i}sica, Universidade Federal do Cear\'{a} (UFC), Campus do Pici, Fortaleza - CE,  60455-760, Brazil.}
\affiliation{Centro de Matématica, Computação e Cognição (CMCC), Universidade Federal do ABC (UFABC), Av. dos Estados 5001, CEP 09210-580, Santo Andr\'{e}, S\~{a}o Paulo, Brazil.}

\begin{abstract}
\noindent \textbf{Abstract:} Modified Gravity Theories (MGTs) are extensions of General Relativity (GR) in its standard formulation. Therefore, within this framework, we will investigate a system composed of a black hole (BH) surrounded by Maxwell-Higgs vortices, forming the BH-vortex system. In the case of linear $f(R)$ gravity is adopted showing the existence of a three-dimensional ring-like BH-vortex system with quantized magnetic flux. Within this system, one notes the BH at $r=0$ and its event horizon at $r= r_0$, while the magnetic vortices are at $r \in (r_0, \infty)$. A remarkable result is the constancy of the Bekenstein-Hawking temperature ($T_H$), regardless of MGTs and vortex parameters. This invariance of $T_H$ suggests that the BH-vortex system reaches thermodynamic stability. Unlike the standard theory of Maxwell-Higgs vortices in flat spacetime, in $f(R)$ gravity, the vortices suffer the influence of the BH's event horizon. This interaction induces perturbations in the magnetic vortex profile, forming cosmological ring-like magnetic structures.
\end{abstract}

\maketitle
\newpage

\section{Introduction} 

Recently, we noted a growing interest in Modified Gravity Theories (MGTs). Generally speaking, these theories arise from the premise that the universe is suffering an accelerated expansion \cite{Tonry,Knop}, which can be explained through corrections to the equations of motion in General Relativity (GR) \cite{GOlmo}. Initially, one explanation for this universal acceleration is the hypothesis that additional contributions from the scalar curvature may exist in the standard formulation of GR, leading to MGTs, e.g.,  $f(R)$ theory \cite{SCapozziello,SMCarroll1}. Specifically, one of the motivations for $f(R)$ gravity is the study of possible new gravitational effects related to late-time cosmic expansion, see Refs. \cite{SDOdin1,SDOdin2,SDOdin3}. These remarks have provided extensive studies on MGTs as promising frameworks for addressing emerging challenges to standard GR \cite{Almeida2}. Among these theories are $f(R)$ \cite{Felice,SCapozziello4} and its variants as $f(R,T)$ gravity \cite{THarkoX}, models involving couplings between curvature and matter \cite{THarko0}, and proposals based on modified geometries \cite{Lavinia,Cai}. Thus, this jungle of MGTs has found significant applications in cosmological and astrophysical contexts \cite{Saito1}.

In several contexts, one adopts widely the MGTs \cite{Zhdanov,Nojiri00,Giacomozzi,Nashed}; however, one knows little about the self-gravitating topological structures\footnote{Topological structures are nonlinear field configurations with stability due to topological properties. These objects arise particularly in field theory and particle physics. Generally speaking, one classifies these structures as kinks, vortices, and monopoles \cite{Vachaspati,Rajaraman,Manton,Vilenkin}.}. A significant class of topological structures are the topological vortices, which emerge from a complex scalar field theory coupled to an Abelian gauge field \cite{Vachaspati,Rajaraman,Manton,Vilenkin}. Initially, Higgs used a massive scalar field in the mechanism that would later bear his name \cite{Higgs}. Subsequently, the investigation of models coupled to an Abelian gauge field was introduced by Nielsen and Olesen \cite{Nielsen}, adopting a flat spacetime. Generally speaking, vortex models preserve the Higgs mechanism and gauge invariance and are particularly relevant once these models are relativistic generalizations of the phenomenological Ginzburg-Landau theory for superconductivity \cite{James}. Thus, the following question naturally arises: would it be possible for a BH-vortex system to emerge in an MGT? We aim to address this question in the development of this letter.

In this letter, we will explore the BH-vortex system within the $f(R)$ gravity. We begin by outlining the general framework for generating the BH-vortex system. Consequently, one investigates the BH and Hawking radiation in $f(R)$ gravity. Finally, the vortices surrounding the BH are analyzed, and our findings are announced.

\section{The BH-vortex system in modified $f(R)$ gravity}

Let us explore the BH and magnetic vortex solutions in modified $f(R)$ gravity. This study is particularly relevant and helps us understand the interaction between the magnetic vortex and the BH. Although this research branch is still relatively unexplored, this study provides us valuable insights into the physical aspects of the BH-vortex system in MGT \cite{Bazeia1,Bazeia2}. This approach also allows us to understand how MGTs, defined as functions of the Ricci scalar ($R$), influence gravitating matter. Furthermore, the Einstein-Hilbert theory corrected by quantum metric fluctuations leads us to consider the Maxwell-Higgs action in $f(R)$ gravity, viz.,
\begin{align}
    \label{1}
    S=\int\, d^3x\, \sqrt{-g}\bigg[-\frac{1}{16\pi}[R+f(R)]+\mathcal{L}_{\mathrm{mat}}\bigg],
\end{align}
where
\begin{align}
    \mathcal{L}_{\mathrm{mat}}=\frac{1}{2}(D_{\mu}\phi)^{\dagger}D^{\mu}\phi-\frac{1}{4}F_{\mu\nu}F^{\mu\nu}-\frac{\lambda}{4}(\vert\phi\vert^2-\nu^2)^2.
\end{align}
Here, $R$ is the Ricci scalar, $\phi$ a complex scalar field, and $F_{\mu\nu}$ the electromagnetic field tensor, defined as $F_{\mu\nu} = \partial_\mu A_\nu - \partial_\nu A_\mu$, where $A_\mu$ is the gauge field. The parameter $\nu$ denotes the Vacuum Expectation Value (VEV). Furthermore, the notation $D_\mu\phi$ is the covariant derivative, which minimally couples the matter field $\phi$ to the gauge field $A_\mu$. One defines the covariant derivative as 
\begin{align}
    \label{2}
    D_{\mu}\phi=\partial_{\mu}\phi+ieA_{\mu}\phi.
\end{align}
Here, $e$ is a minimal coupling between the gauge and matter fields\footnote{For simplicity, let us adopt the natural units system, which leads us to assume $\hbar=c=e=1$.}.

To obtain static and topological structures in three-dimensional spacetime \cite{Edery}, let us adopt the metric 
\begin{align}\label{3}
    ds^2=-h(r)\,dt^2+\frac{1}{h(r)}dr^2+r^2\, d\theta^2,
\end{align}
where $h(r)$ is the metric function, $r$ and $\theta$ are, respectively, the radial and angular variables\footnote{For more details on this metric, see Refs. \cite{Edery}. In Refs. \cite{Edery}, the authors used this metric to examine the vortex solution in Einstein's gravity.}.

Magnetic topological vortices are gauge-invariant and rotationally symmetric structures \cite{Vachaspati,Rajaraman,Manton,Vilenkin}. Thus, one defines the gauge field as
\begin{align}
    \label{4}
    \textbf{A}_i(\textbf{r})=-\varepsilon_{ij}\hat{x}^{j}\frac{a(r)}{er},
\end{align}
which leads us to structures with magnetic flux ($\Phi_B$) determined by
\begin{align}
    \label{flux}
    \Phi_B=\frac{2\pi}{e}[a(0)-a(\infty)].
\end{align}

Let us concentrate on the study of topological vortices. To fulfill this purpose, we will adopt the topological conditions \cite{Vachaspati,Rajaraman,Manton}
\begin{align}
    \label{cond}
    a(0)=0 \, \, \, \, \, \text{and} \, \, \, \, \, a(\infty)=-\beta \, \, \, \, \text{with} \, \, \, \, \, \beta\in \mathds{Z}^{+}.
\end{align}

Assuming the topological condition announced in Eq. \eqref{cond}, one obtains magnetic vortices with magnetic flux given by
\begin{align}
    \Phi_B=\frac{2\pi \beta}{e}.
\end{align}
Accordingly, one concludes that the topological vortices radiate a quantized magnetic flux.

For the existence of magnetic vortices, the $U(1)$ symmetry must, a priori, be omnipresent. Thus, as outlined in prior research \cite{Vachaspati,Rajaraman,Manton,Nielsen}, let us adopt the \textit{ansatz} 
\begin{align}
    \label{5}
    \phi(r,\theta)=g(r)\text{e}^{in\theta},
\end{align}
where $n \in \mathbb{Z}^{+}$. Furthermore, $n$ is the winding number
\footnote{The winding number is a quantity concerning the complex scalar field associated with the vortex. Supposing a vortex at $r=\bar{r}_0$, one can interpret the winding number as the amount of turns the function (or field) wraps around the vortex (at $r=\bar{r}_0$) along a closed trajectory. Therefore, the winding number plays a significant role, allowing us to understand the topological aspects of the vortex, e.g., the magnetic field \cite{Neu}.} concerning the vortex \cite{FLima1,FLima2,FLima3,FLima4}. The functions $g(r)$ and $a(r)$ are, respectively, the field variables for the matter ($\phi$) and gauge ($A_\mu$) sector.

Now, allow us to investigate the BH-vortex system and examine the resulting metric signature in $f(R)$ gravity. To attain our target, we will consider the equation of motion derived from varying the action \eqref{1} concerning the metric. This variation leads to 
\begin{align}
    \label{7}
    T_{\mu\nu}=&R_{\mu\nu}(1+f_R)-\frac{1}{2}g_{\mu\nu}[R+f(R)]+(\nabla_\mu\nabla_\nu-g_{\mu\nu}\nabla_{\beta}\nabla^{\beta})f_R, 
\end{align}
where $f_R$ is the total derivative of the function $f(R)$ concerning $R$.

Additionally, one defines the stress-energy tensor\footnote{This quantity will be essential in the following sections to examine the BH-vortex metric and the system's energy.} as
\begin{align}\label{8}
    T_{\mu\nu}=-\frac{2}{\sqrt{-g}}\frac{\delta S}{\delta g^{\mu\nu}}.
\end{align}

To determine the profile of metric functions in gravity $f(R)$ theory, we will focus our study on cases where the effects of the presence of vortices are minimally perceptible. In other words, let us analyze regions outside the BH-Vortice system, i.e., sectors where the effects are negligible. In this regime, one can assume $T_{\mu\mu} = 0$ [Eq. \eqref{8}]. This simplification allows us  to obtain using Eq. \eqref{7}, three independent differential equations, corresponding to the components $tt$, $rr$, and $\theta\theta$, i.e.,
\begin{align}
     \label{9}
    &2rh(r)f''_{R}+[rh'(r)+2h(r)]f'_{R}+[rh''(r)+h'(r)]f_{R}-h'(r)=-rf(R);\\ \label{10}
    &[rh'(r)+2h(r)]f'_R+[rh''(r)+h'(r)]f_R-h'(r)=-rf(R);\\ \label{11}
    &2rh(r)f''_{R}+2rh'(r)f'_R+2h'(r)f_{R}-rh''(r)=-rf(R),
\end{align}
which leads us to 
\begin{align}
    \label{Eq13}
    2rh'(r)f'_R+2h'(r)f_{R}-rh''(r)+rf(R)=0. 
\end{align}
Additionally, through algebraic manipulations of Eqs. \eqref{9} and \eqref{10}, we concluded that $f''_R = 0$. It is important to highlight that the prime notation is the derivative concerning the radial variable $r$.

Now, let us assume the case in which the metric satisfies the condition that the effects of the structure are minimally perceptible outside the BH-vortex system. Seen from this standpoint, the modified gravity must take the form $f(R)=\alpha R$, which allows for expressing
\begin{align}
    \label{12}
    R=-\frac{2h'(r)}{r}-h''(r)
\end{align} 
and 
\begin{align}
    f(R)=-\alpha\left[\frac{2h'(r)}{r}+h''(r)\right]
\end{align}
which leads us to
\begin{align}\label{alphaFR}
    (1+\alpha) r h''(r)=0 \hspace{0.5cm} \text{with} \hspace{0.5cm} \alpha\neq -1.
\end{align}

Consequently, Eq. \eqref{alphaFR} provides us with the solution
\begin{align}\label{Eqh}
    h(r)=r\mp r_0.
\end{align}
Our purpose is to obtain BH-like compact objects. In pursuit of this, we consider only physically admissible solutions [Eq. \eqref{Eqh}] from the theory, i.e., $h(r)=r-r_0$\footnote{The parameter $r_0$ may take arbitrary values. However, it is convenient to keep $r_0$ small once it is a comparison of the vortex matter and the horizontal event. That convenience is because minimal variations in $r_0$ induce abrupt changes in the magnetic vortices.}. Indeed, the function $h(r)=r-r_0$ represents a simple and linear form of a curvilinear coordinate function in three-dimensional spacetime. In this framework, one can interpret the parameter $r_0$ as the BH mass. Thus, this function constitutes a modification of the radial term of the Schwarzschild-like metric. By analogy with the Schwarzschild-like BH, this metric function profile implies that the temporal term assumes negative values for $r<r_0$, suggesting the existence of an event horizon at $r=r_0$. Furthermore, in the limit $r \to r_0$, the temporal contribution vanishes, ensuring an essential feature of an event horizon \cite{Carroll} \footnote{It is worth emphasizing that, in the limit $\alpha\to 0$, the action \eqref{1} reduces to the standard (na\"{i}ve) formulation of the theory, viz., $S=\int\, d^{3}x\, \sqrt{-g}\left[-\frac{1}{16\pi}R+\mathcal{L}_{\text{matter}}\right]$. In this regime, the Einstein field equations of general relativity are recovered, i.e., $R_{\mu\nu} - \frac{1}{2}g_{\mu\nu}R = T_{\mu\nu}$, which admit recovering the Schwarzschild solution as a particular case. Alternatively, by parametrizing the theory through the inclusion of a $f(R)$-term arising from quantum fluctuations of the metric, viz., $f(R) = \alpha R$, one obtains a Schwarzschild-like theory, in which we reorganize all relevant parameters in the constant $r_0$. Furthermore, it is noteworthy that Eq. \eqref{Eq13} reduces to $h''(r)=0$ in the limit of the standard theory, precisely matching the result found in the classical Schwarzschild solution. For further details, see Refs. \cite{Carroll}.}.

\begin{figure}[!ht]
    \centering
    \includegraphics[width=0.55\linewidth]{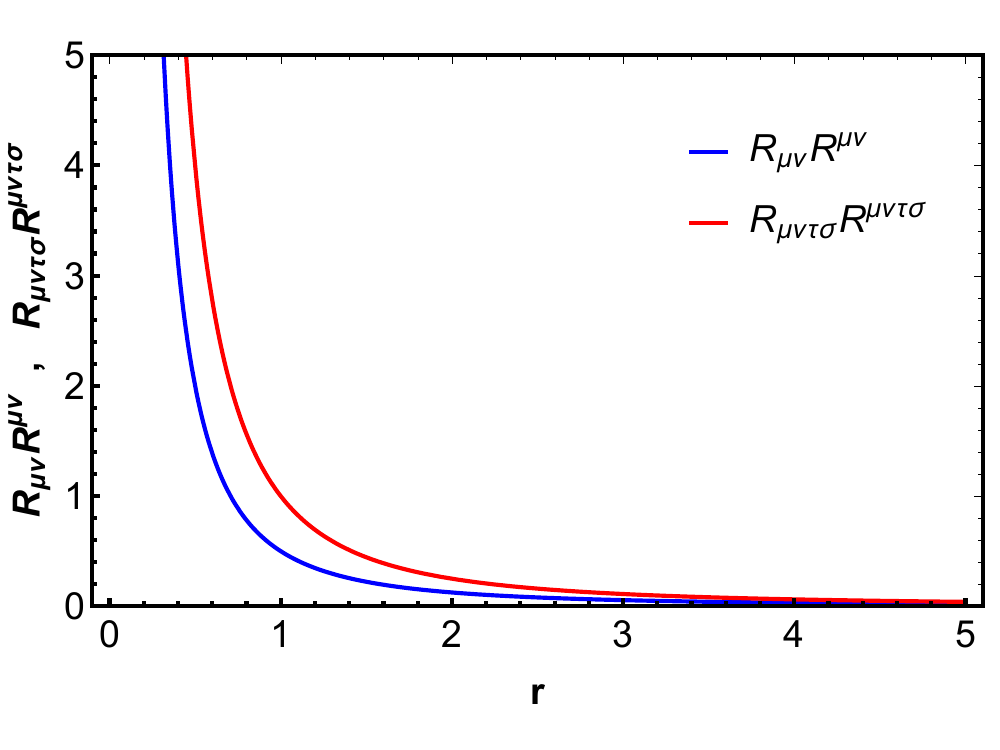}
    \caption{Quadratic invariant of the Ricci tensor (blue curve) and the Kretschmann scalar (red curve) versus radial distance. This plot is valid for any value of $\alpha\neq-1$.}
    \label{FigCurv}
\end{figure}

Once we have the emergence of a BH-like compact object, it becomes necessary to evaluate the stability of this cosmological object. For this assessment, let us adopt the metric function solution $h(r)=r-r_0$ [Eq. \eqref{Eqh}] and investigate the quadratic invariant of the Ricci tensor and the Kretschmann scalar. Through direct calculation for the metric \eqref{3}, the quadratic invariant of the Ricci tensor is 
\begin{align}
    R^{\mu\nu}R_{\mu\nu}=\frac{1}{2r^2}h'(r)[h'(r)+rh''(r)]=\frac{1}{2r^2}.
\end{align}
Additionally, one defines the Kretschmann scalar ($K$) by the quadratic invariant of the Riemann curvature tensor ($R_{\mu\nu\tau\sigma}$). Thus, for our conjecture, the Kretschmann scalar is
\begin{align}
    K=R^{\mu\nu\tau\sigma}R_{\mu\nu\tau\sigma}=\frac{h'(r)^2}{r^2}=\frac{1}{r^2}.
\end{align}
Analyzing the Kretschmann scalar (see Fig. \ref{FigCurv}), we noted the existence of a singularity at the origin. Asymptotically, the gravitational field approaches zero curvature, indicating that the BH (and the vortex) does not affect the curvature. Besides, the results presented in Fig. \ref{FigCurv} ensure a BH at $r=0$.
 
Considering the metric function profile described in Eq. \eqref{Eqh}, one concludes that the $f[R(r)]$ gravity produced by the correction is 
\begin{align}\label{Gfr}
    f[R(r)]=-\frac{2\alpha}{r}.
\end{align}
When considering a linear function $f(R)$, one notes that the contribution arising from the quantum correction of the metric modifies the geometric sector, enabling adjustments to the effective Newtonian constant. This modification of the gravitational constant intensifies the deformations in the matter sector, altering the vortex's matter configuration of the BH-vortex system. Consequently, this modification induces changes in the radiation emitted by the system as part of the matter collapses into the black hole. Therefore, this process affects phenomena on local scales, i.e., near the BH-vortex system \footnote{One can note this deformed behavior in Figs. \ref{fig2}(a) and \ref{fig2}(b). While in Einstein's classical theory, the vortices exhibit profiles resembling kinks, in the $f(R)$ gravity theory announced in this work, deformations occur in the matter sector, specifically in the complex scalar and gauge fields. For further details, see Refs. \cite{Edery}.} .

\subsubsection{On the Bekenstein--Hawking temperature} 

In our system, we have a three-dimensional BH emitting Hawking radiation. Thus, it is essential to study this radiation. To fulfill this, let us use the Hamilton-Jacobi formalism through the tunneling approach to examine the thermodynamics concerning the BH \cite{Srinivasan,Angheben,Kerner1,Mitra,Akhmedov}. This approach is fruitful once it allows us to calculate the Hawking temperature for the black hole surrounded by magnetic vortices, using the metric function presented in Eq. \eqref{Eqh}.

The fundamental conception of the tunneling method involves calculating the probability of particles interacting near the event horizon and escaping through a quantum tunneling process. This interpretation becomes feasible when Hawking radiation is understood as an intrinsic emission process of the black hole. In our system, one can interpret this process as an emission resulting from the spontaneous creation of particle pairs near the event horizon. Thus, this emission is driven by the matter sector, which undergoes decay into the black hole. Consequently, particles with negative energy remain inside the black hole, contributing to its mass decrease. At the same time, particles with positive energy escape the horizon through a ``tunnel'' toward infinity, influencing the energetic changes of the vortex. These changes perturb the massive scalar field solutions and modify the matter sector near the event horizon.

We are ready to study the tunneling probability related to the BH temperature. A significant advantage of this approach for studying the thermodynamics of BH is that thermodynamic properties are intrinsically bounded to the geometry, allowing for its broad application to various spacetime manifolds \cite{Jiang,Kerner2,Ma,Gomes}. To perform this investigation, one knows that, near the BH event horizon, only the temporal and radial terms of the metric remain relevant, while the angular part is redshifted \cite{Brito}. Thus, for our metric \eqref{3}, one obtains 
\begin{align}\label{EqDs}
    ds^2=-(r-r_0)dt^2+\frac{1}{r-r_0}dr^2.
\end{align}

Let us apply a perturbation to the massive scalar sector $\phi$ around the BH background. That allows us to obtain
\begin{align}\label{KG}
    \nabla^{\mu}\nabla_{\mu}\phi-m^2\phi=0.
\end{align}
This expression corresponds to the Klein-Gordon equation in natural unit system \cite{KleinG}, where $m$ is the mass of the field $\phi$. By considering the decomposition of the Klein-Gordon equation \eqref{KG} into spherical harmonics, we come to
\begin{align}\nonumber
    -\frac{\partial^2\phi}{\partial t^2}+(r-r_0)^2\frac{\partial^2\phi}{\partial r^2}+\frac{1}{2}\frac{\partial}{\partial r}(r-r_0)^2\frac{\partial\phi}{\partial r}-m^2 (r-r_0)\phi=0&.
\end{align}

Assuming that the particle creation process in the BH is semiclassical in the BH background, the WKB approximation informs us that the \textit{ansatz} for the field $\phi$ \cite{Srinivasan,Angheben,Kerner1,Mitra,Akhmedov} is
\begin{align}
    \phi(t,r)=\text{e}^{\Theta(t,r)},
\end{align}
which leads us to
\begin{align}\label{Harm}
    \left(\frac{\partial\Theta}{\partial t}\right)^2-(r-r_0)^2\left(\frac{\partial \Theta}{\partial r}\right)^2-m^2(r-r_0)=0,
\end{align}
such that the particle-like solutions are
\begin{align}\label{RTheta}
    \Theta(t,r)=-\omega t+W(r),
\end{align}
where $\omega$ is a constant representing the energy of the emitted radiation. 

By substituting Eq. \eqref{RTheta} into Eq. \eqref{Harm}, boil down to the function $W(r)$, i.e.,
\begin{align}\label{EqW}
    W(r)=\pm\int\,\frac{\sqrt{\omega^2-m^2(\bar{r}-r_0)}}{\bar{r}-r_0}\,d\bar{r}.
\end{align}
The positive sign is associated with outgoing particles, while the negative solution corresponds to incoming particles. Hence, we will focus on the outgoing particles responsible for emitting radiation as they cross the event horizon. Thus, solving Eq. \eqref{EqW} and considering only the outgoing particles (i.e., the positive sign), one concludes that 
\begin{align}
    W(r)=2\pi i\omega+\text{(real contribution)},
\end{align}
As a result, the probability of a particle escaping the BH via the tunneling process is    $\Gamma\sim\text{exp}(-2im\Theta)=\text{exp}(-4\pi\omega).$ Recalling that the tunneling probability is related to the Boltzmann factor $\text{exp}(-\omega/T)$ \cite{Pathria}, it follows that the Hawking temperature will be
\begin{align}
    T_H=\frac{\omega}{2im\Theta}=\frac{1}{4\pi}=0.0795.
\end{align}
Therefore, one concludes that for the BH with an event horizon at $r=r_0$, the Bekenstein-Hawking temperature ($T_H$) remains constant, independent of the $f(R)$ gravity ($\alpha$) parameter or the vortex ($\lambda$ and $\nu$). This constancy in $T_H$ suggests that the BH-vortex system is stable and maintains thermodynamically stable.

\section{The topological vortex}

We will focus on the study of magnetic vortices. To accomplish this, let us analyze the equations of motion within the context of modified $f(R)$ gravity (\ref{Gfr}). In this framework, one notes that the equations of motion derived through the variation of the action concerning the scalar field, the gauge field, and the metric are, respectively,
\begin{align} \label{mov1}
    &r(r-r_0)g''(r)+(2r-r_0)g'(r)+\lambda (g^2-\nu^2)g(r)-[n-a(r)]^2 g(r)=0;&\\ \label{mov2}
    &-r(r-r_0)a''(r)-r_0 a'(r)+r[n-a(r)]g(r)^2=0;\\ \label{mov3}
    &g'(r)=\frac{1}{r}a'(r).
\end{align}
Note that the set of linearly independent equations [\eqref{mov1}-\eqref{mov3}] is also invariant concerning the parameter $\alpha$. This behavior arises because cosmological magnetic vortices are classical objects not influenced by small quantum fluctuations originating from the quantum correction to the metric. Conversely, these vortices are significantly affected by the BH radius. That occurs because, as the event horizon expands, the vortex matter begins to collapse into the BH, causing increasing distortions in the magnetic vortex. We expose a representation of the collapse of the magnetic vortex matter into the BH.

By adopting the equations of motion that describe the topological vortices [Eqs. \eqref{mov1}-\eqref{mov3}], we are ready to investigate the numerical solution of these vortices, taking into account the topological conditions expressed in \eqref{cond} along with
\begin{align}\label{CondTop00}
    g(0)=0 \hspace{0.5cm} \text{and} \hspace{0.5cm} g(\infty)=\nu.
\end{align}
Here, $\nu$ is the VEV. For our analysis, we adopt, for simplicity, $\lambda=\beta=\nu=n=1$ \footnote{Once $\nu$ describes the asymptotic value of the complex scalar field, i.e., $g(\infty) = \nu$, its modification only affects the value of $g(\infty)$, as announced in Eq. \eqref{CondTop00}. Meanwhile, the parameter $\lambda$ becomes the topological structures more (or less) localized. Similarly, the parameter $\beta$ modifies the asymptotic value of the gauge field, i.e., $a(\infty)$ [see Eq. \eqref{cond}]. Thus, for convenience, when studying topological vortex configurations, it is common to assume these parameters to be unitary, i.e., $\nu=\lambda=\beta=1$, once they do not produce any physical repercussions. For further details, see Ref. \cite{FLima3}.}. Additionally, to examine the numerical solutions of the system of equations [\eqref{mov1}-\eqref{mov3}], we will use the numerical interpolation method, which allows us to estimate the solutions for the differential equations [Eqs. \eqref{mov1}-\eqref{mov3}]. That will provide us with the solutions for the field variables $g(r)$ and $a(r)$ that satisfy the Eqs. [\eqref{mov1}-\eqref{mov3}]. Generally speaking, we apply the numerical interpolation method and discretize the function's domain. For instance, by considering the differential equations [\eqref{mov1}-\eqref{mov3}] defined in the range of the radial variable $r$, this range is subdivided into discrete steps. Thus, we discretize the position of the system by dividing the position range into two hundred thousand discrete points. Therefore, the solutions of the equations are investigated within the range $[0, 200]$, using the discretization $r_1,\, r_2,\,\dots,\, r_{200000}$, where $r_1 = 0$ and $r_{200000}=200\,$\footnote{For further details on the numerical method, see Refs. \cite{Scarb,EBurden}.}.

Using the numerical interpolation approach \cite{Scarb,EBurden}, let us investigate the numerical vortex solutions, i.e., the solutions for the field variables $g(r)$ and $a(r)$. We display the numerical solutions for the field variables $g(r)$ and $a(r)$, respectively, in Figs. \ref{fig1}(a) and \ref{fig1}(b).
\begin{figure}[!ht]
  \centering
  \subfigure[Field variable of the matter sector.]{\includegraphics[width=0.5\linewidth]{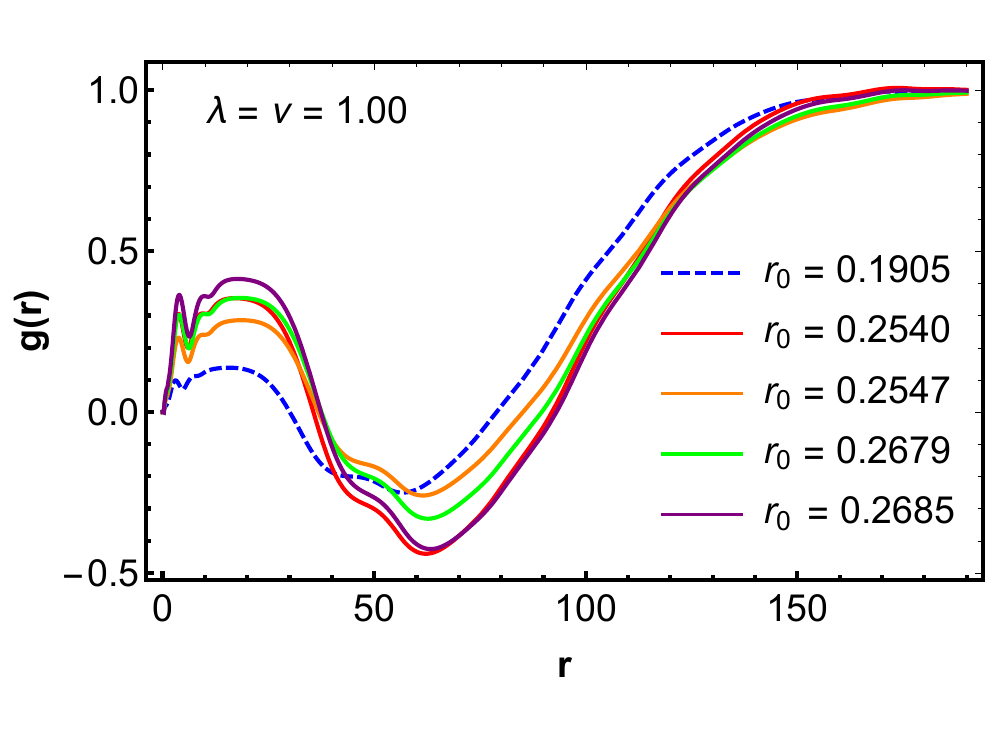}}\hfill
  \subfigure[Field variable of the  gauge sector.]{\includegraphics[width=0.5\linewidth]{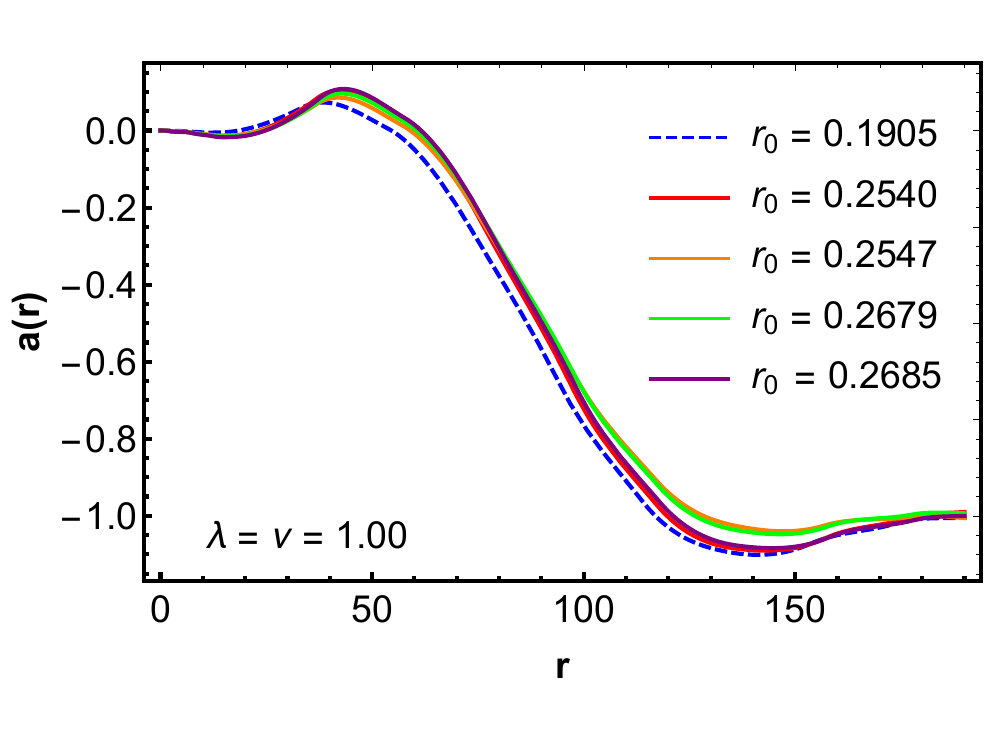}}\hfill
  \caption{The numerical vortex solutions.}  \label{fig2}
\end{figure}

From the analysis of the numerical vortex solutions in $f(R)$ gravity with the varying event horizon of the BH, one can note that the vortices are sensitive to the presence of the compact cosmological object. This compact structure generates perturbations in the vortex core, resulting in the formation of vortex structures with the $g(r)$ sector experiencing abrupt changes near the event horizon, see Fig. \ref{fig2}(a). Indeed, this behavior is a consequence of the collapse of the magnetic vortex matter into the BH.

Analyzing the quadratic invariant of the Ricci scalar and the Kretschmann scalar, we noted the BH core at $r=0$. The presence of the BH promotes a local minimum near the event horizon, followed by an absolute maximum in the gauge sector before rapidly stabilizing to the asymptotic value $a(r\to\infty)\to -\beta$, as shown in Fig. \ref{fig2}(b).

\begin{figure}[!ht]
  \centering
  \subfigure[Magnetic field vs. $r$.]{\includegraphics[width=0.5\linewidth]{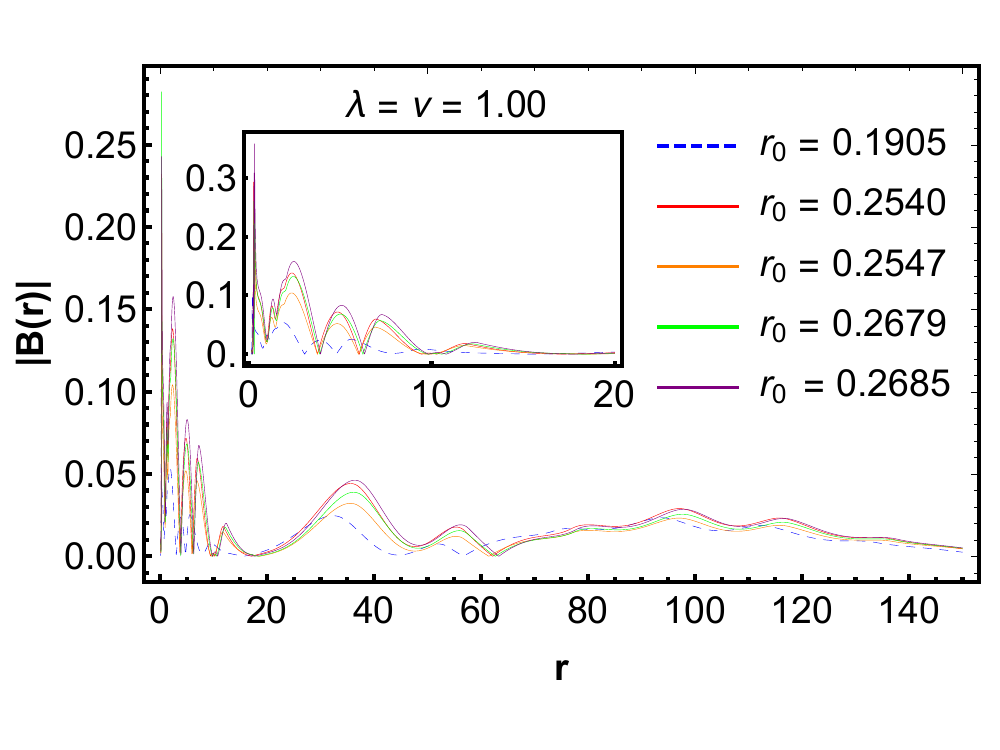}}\hfill
  \subfigure[Energy density vs. $r$.]{\includegraphics[width=0.5\linewidth]{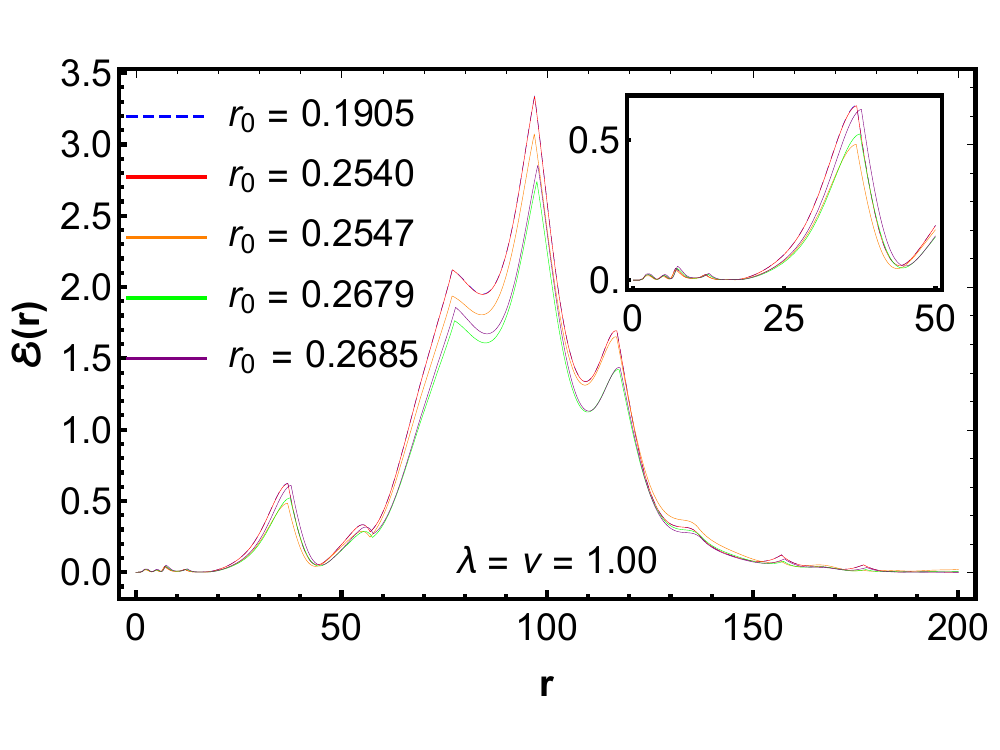}}\hfill
  \caption{The numerical vortex solutions.}  \label{fig3}
\end{figure}

Finally, evaluating the magnetic field\footnote{One defines the intensity of the magnetic field of the vortex as $B = -F_{21} \equiv \frac{a'(r)}{er}$ \cite{Weinberg}.} and energy density\footnote{The energy density of the BH-vortex system is the $00$-component of the stress-energy tensor [$00$-component of Eq. \eqref{8}] \cite{Vachaspati,Rajaraman,Manton}.} of the BH-vortex system, we noted the existence of various magnetic oscillations near the event horizon of the BH, with energy density localized in the vortex. Naturally, the magnetic vortices formed in f(R) gravity are of the ring-like structures. We exposed the ring-like vortices in Figs. \ref{fig3}(a) and \ref{fig3}(b).

\section{Summary and conclusion}

We employed Heisenberg's non-perturbative approach to promote the three-dimensional Einstein-Hilbert theory to a modified $f(R)$ gravity. In this conjecture, we studied a BH-vortex system. Within this framework, one shows the existence of a static three-dimensional black hole surrounded by magnetic ring-like vortices responsible for generating quantized magnetic flux and an anisotropic energy density near the event horizon, i.e., $r = r_0$. An analytical analysis of the quadratic invariant of the Ricci scalar ($R_{\mu\nu}R^{\mu\nu}$) and the Kretschmann scalar ($R_{\mu\nu\tau\sigma}R^{\mu\nu\tau\sigma}$), announced a BH at $r=0$. Meanwhile, the vortices surround the black hole with their innermost rings close to $r \approx r_0$. By examining the thermodynamics of the black hole originating from vortex matter collapsing into the black hole, we found that the Bekenstein-Hawking temperature remains constant, independent of the parameters governing the MGT and the magnetic vortex. This result suggests that the BH-vortex system is thermodynamically stable. Lastly, the vortices are directly influenced by the event horizon, causing perturbations in the magnetic profile of the vortices and leading to the formation of cosmological structures with magnetic ring-like profiles.

Starting from the premise that outside of the BH-vortex system $T_{\mu\mu}=0$, one obtains that $h(r)=r-r_0$, with $f(R)=-2\alpha r^{-1}$, where $\alpha$ is the quantum fluctuation of the metric. Thus, we can reinterpret $r_0$ as a parameter associated with the BH mass. Naturally, this function represents a modification of the radial term in the Schwarzschild metric. By analogy to the Schwarzschild-like black hole, the profile of this metric function implies that the temporal term takes negative values for $r<r_0$, indicating the existence of an event horizon at $r=r_0$. Additionally, when $r \to r_0$, the temporal factor vanishes, which characterizes an essential feature of an event horizon. These results corroborate with the results for the quadratic invariant of the Ricci scalar and the Kretschmann scalar, which ensure the existence of a black hole at $r = 0$, with an event horizon at $r=r_0$.

Examining the magnetic vortices around the three-dimensional BH, we noted that the field variables [$g(r)$ and $a(r)$] describing the vortex are highly sensitive to the presence of the compact object (BH) at $r=0$ and their dynamics are altered with alters of the event horizon ($r_0$). Consequently, these vortices experience significant perturbations near the event horizon. This behavior naturally arises from the collapse of matter present in the magnetic vortex into the black hole. We concluded that the magnetic field of the vortex exhibits a ring-like profile, showing various magnetic oscillations near the event horizon with energy density is anisotropically distributed.

\section{ACKNOWLEDGMENT}
The authors are grateful to the Conselho Nacional de Desenvolvimento Científico e Tecnológico (CNPq). F. C. E. Lima is supported, respectively, for grants No. 171048/2023-7 (CNPq/PDJ) and 309553/2021-0 (FAPESP/PQ).

\end{document}